\documentclass[prd,aps,floats]{revtex4}
\usepackage{amsmath}
\usepackage{amssymb}

\numberwithin{equation}{section}

\usepackage{epsfig}

\usepackage{setspace}
\def\beq{\begin{equation}}
\def\eeq{\end{equation}}
\def\ber{\begin{eqnarray}}
\def\eer{\end{eqnarray}}

\def\atridot{\stackrel{...}{a}}
\def \lleq {\lower0.9ex\hbox{ $\buildrel < \over \sim$} ~}
\def \ggeq {\lower0.9ex\hbox{ $\buildrel > \over \sim$} ~}
\def\apj{{Astroph.\@ J.\ }}
\def\mn{{Mon.\@ Not.\@ Roy.\@ Ast.\@ Soc.\ }}

\def\prl{{Phys.\@ Rev.\@ Lett.\ }}
\def\prd{{Phys.\@ Rev.\@ D\ }}

\def \jetpl {JETP Lett.\ }
\def\etal{{\it et al.}}

\begin{document}


\title{A new null diagnostic customized for reconstructing the
properties of dark energy from BAO data}

\author{Arman Shafieloo$^{a,b,c}$, Varun Sahni$^d$, and Alexei A. Starobinsky$^{e,f}$}
\affiliation{$^a$  Asia Pacific Center for Theoretical Physics, Pohang, Gyeongbuk 790-784, Korea}
\affiliation{$^b$ Department of Physics, POSTECH, Pohang, Gyeongbuk 790-784, Korea}
\affiliation{$^c$ Institute for the Early Universe WCU, Ehwa Womans University,
Seoul, 120-720, Korea }
\affiliation{$^d$ Inter-University Centre for Astronomy and Astrophysics,
Post Bag 4, Ganeshkhind, Pune 411~007, India}
\affiliation{$^e$ Landau Institute for Theoretical Physics RAS,
Moscow 119334, Russia}
\affiliation{$^f$ Research Center for the Early Universe (RESCEU),  
Graduate School of Science, The University of Tokyo, Tokyo 113-0033, Japan} 

\thispagestyle{empty}

\sloppy

\begin{abstract}
Baryon Acoustic Oscillations (BAO) provide an important standard ruler which can be
used to probe the recent expansion history of our universe. We show how a simple
extension of the $Om$ diagnostic, which we call $Om3$, can combine standard
ruler information from BAO with standard candle information from type Ia 
supernovae (SNIa) to yield a powerful novel null diagnostic of the cosmological constant
hypothesis. A unique feature of $Om3$ is that it requires
{\em minimal cosmological
assumptions} since its determination  does not rely upon prior knowledge of either
the current value of the matter density, $\Omega_{0m}$ and 
the Hubble constant $H_0$, or the distance to the last scattering surface. 
Observational uncertainties in these 
quantities therefore do not affect the reconstruction of $Om3$.
We reconstruct $Om3$ using the Union 2.1 SNIa data set and BAO data from 
SDSS, WiggleZ and 6dFGS. Our results are consistent with dark energy being
the cosmological constant. We show how $Om$ and $Om3$
can be used to obtain accurate model independent constraints on the properties of dark energy
from future data sets such as BigBOSS.

\end{abstract}

\maketitle


\bigskip

\section{Introduction}
\label{sec:intro}
The expansion of the universe appears to have undergone a dramatic change in
its recent past with most observations suggestive of the fact that
cosmic expansion began to accelerate when the scale factor $a(t)$ of the Universe was 
about $0.6$ of its present value, i.e. at the redshift $z\sim 0.7$. Although cosmic acceleration is 
a dramatic phenomenon, evidence for it is indirect and stems from data observed along our past light 
cone and involving either standard candles (SNIa) or standard rulers (BAO, CMB) 
\cite{observations,Union21,percival,wigglez1,wigglez2,6dF}.

The raison d'etre behind cosmic acceleration remains unknown and the fact that it might 
signal the need for new physics is partly responsible for the high levels of activity
marking this field.
Possible drivers of acceleration range from the Einstein's early suggestion of the 
cosmological constant \cite{einstein} to more elaborate constructions collectively called Dark Energy (DE)
which may equally well be based on the introduction of a new physical field (physical DE) or on
modifying the Einstein General Relativity (GR) (geometrical DE), or both \cite{DE_review,ss06}.
To distinguish between these widely varying alternatives, 
model independent diagnostic tools capable of differentiating different 
classes of DE models are of great help \cite{ss06}.

With this in mind, 
we introduced the $Om$ diagnostic \cite{Om,zunkel_clarkson} which successfully distinguishes 
evolving DE from the cosmological constant on the basis of observations of
the expansion history $H(z)$. The two-point $Om$ diagnostic can be defined as 
follows 
\beq
Om(z_2;z_1) = \frac{h^2(z_2)-h^2(z_1)}{(1+z_2)^3 - (1+z_1)^3}, ~~h(z) = H(z)/H_0
\label{eq:om}
\eeq
so that \cite{Om}
\beq
Om(z;0) \equiv Om(z) = \frac{h^2(z)-1}{(1+z)^3 - 1}~.
\eeq
For a spatially flat universe, $H(z)$ can be recovered from the luminosity distance
via a single differentiation
\cite{st98,HT99,chiba99,saini00,chiba00}
\beq\label{eq:H}
H(z)=\left[{d\over dz}\left({D_L(z)\over 1+z}\right)\right]^{-1}~,
\eeq
which makes $H(z)$ more robustly determined than the equation of state
\beq\label{eq:state}
w(x) =
\frac{(2 x /3) \ d \ {\rm ln}H \ / \ dx - 1}{1 \ - \ (H_0/H)^2
\Omega_{0m} \ x^3}\,\,, ~~x = 1+z~,
\eeq
which involves knowing the second derivative of $D_L$ as well as the cosmological
matter density $\Omega_{0m}$. Thus observational uncertainties in $D_L$ and
$\Omega_{0m}$ muddy the reconstruction of $w(z)$ to a much greater extent
than $H(z)$ and $Om(z)$ \cite{Om,zunkel_clarkson}. The expansion history, $H(z)$, can also
be determined directly from ages of passively evolving galaxies \cite{simon}, 
the time drift of the cosmological redshift
\cite{sandage}, 
radial BAO's \cite{seo-eisenstein,gaztanaga,review} or from a combination of the Alcock-Paczynski test
and galaxy redshift space distortions \cite{blake}.
This opens up the possibility of determining $Om(z)$ from a combination of data sets
involving the low redshift universe ($z \lleq {\rm few}$), precisely where
cosmic acceleration seems to have originated !

\begin{figure*}[ht]
\centerline{ \psfig{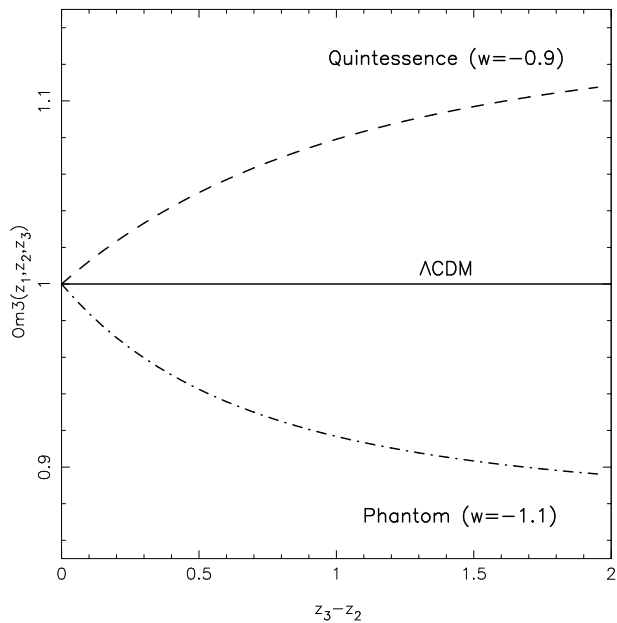} }
\bigskip
\caption{\small
The $Om3$ diagnostic (\ref{eq:om3})
 is shown as a function of separation, $z_3-z_2$, for
$z_1 = 0.2, z_2 = 0.35$ and $\Omega_{0m} = 0.27$. This diagnostic is clearly able to
discriminate quite well between evolving DE models and $\Lambda$CDM,
since, for $\Lambda$CDM the value of $Om3$ stays pegged at unity, while for 
Quintessence/Phantom $Om3$ rapidly evolves to larger/smaller values as  
the separation $z_3-z_2$ increases.
}
\label{fig:eos}
\end{figure*}

The possibility of using $Om$ as a {\em null diagnostic} follows from 
the fact that, for the cosmological constant 
\beq
Om(z_2;z_1) = \Omega_{0m}~.
\label{eq:LambdaOm}
\eeq
In other words, the value of $Om$ is {\em redshift independent} for
the cosmological constant, while for other models of cosmic acceleration $Om(z)$
is {\underline {redshift dependent}}.
This important property of $Om$ can be used to considerable advantage by
defining the {\em difference diagnostic} \cite{Om}
\beq
Om_{\rm diff}(z_1,z_2) := Om(z_1) - Om(z_2)
\label{eq:om2}
\eeq
and the {\em ratio diagnostic}
\beq
Om_{\rm ratio}(z_1,z_2,z_3,z_4) := \frac{Om(z_2;z_1)}{Om(z_4;z_3)}~. 
\label{eq:om4}
\eeq
From (\ref{eq:LambdaOm}) one immediately finds that the following equalities 
must hold for
 the cosmological constant 
\beq
Om_{\rm diff}(z_1,z_2) = 0, ~~ Om_{\rm ratio}(z_1,z_2,z_3,z_4) = 1~.
\label{eq:defn1}
\eeq
A departure of $Om_{\rm ratio}$ from unity 
therefore serves as a `smoking gun' or 
{\em null test} for concordance cosmology: $\Lambda$CDM.

If any two of the four redshifts in $Om_{\rm ratio}$ are identical
then $Om_{\rm ratio}$ reduces to the {\em 3 point diagnostic} $Om3$
\beq
Om_{\rm ratio}(z_1,z_2,z_1,z_3) := Om3(z_1,z_2,z_3)
= \frac{Om(z_2;z_1)}{Om(z_3;z_1)}~, 
\label{eq:om3}
\eeq
where $Om3 = 1$ for $\Lambda$CDM.$^1$\footnotetext[1]{Note a passing similarity
between the diagnostic pair $\lbrace Om_{\rm ratio}, Om_{\rm diff}\rbrace$ and the Statefinder
diagnostic \cite{statefinder} $\lbrace r,s \rbrace$, where $r = \atridot/a H^3$,
$s= (r-1)/3(q-1/2)$. In both cases, for the $\Lambda$CDM model one finds
 $\lbrace Om_{\rm ratio}, Om_{\rm diff}\rbrace$ = $\lbrace 1, 0\rbrace$
and $\lbrace r,s \rbrace$ = $\lbrace 1, 0\rbrace$. 
However, unlike the statefinder pair, the values of $Om_{\rm diff}$ \& $Om_{\rm ratio}$ are strongly
correlated, as seen from (\ref{eq:om2}) and (\ref{eq:om3}).
Additionally, neither of 
$Om_{\rm diff}, Om_{\rm ratio}$ 
is able to distinguish steady state cosmology (SS) 
($\Omega_\Lambda = 1, \Omega_{0m} = 0$),
from SCDM ($\Omega_\Lambda = 0, \Omega_{0m} = 1$) while the $Om$ diagnostic and
the Statefinders are more
successful on this score, since $Om(z) = 0$ for SS while $Om(z) = 1$ for SCDM;
similarly $\lbrace r,s\rbrace = \lbrace 1,0 \rbrace$ for SS while $\lbrace r,s\rbrace = \lbrace 1,1 \rbrace$ for SCDM (see also~\cite{maryam_varun}).
}
(See \cite{shaf_clark,campo,alam} for other discussions of the $Om$ diagnostic.)

In this paper we focus on $Om3$ and show that the main difference between $Om3$ and $Om$ 
is that these two diagnostics test the cosmological constant hypothesis using different cosmological observables. Whereas $Om$ is a powerful null diagnostic of the cosmological constant as it requires only a prior knowledge 
of the expansion history, $h(z)$, one should note that deriving $h(z)$ directly from cosmological observables in a purely model independent and non-parametric manner is not always a simple task. Although 
 supernovae data do allow one to reconstruct $h(z)$ in a model independent manner 
\cite{smooth}, deriving $h(z)$ directly from large scale structure data seems to be 
considerably more difficult (see~\cite{shaf_clark} for detailed discussions). In contrast to $Om$, the $Om3$ diagnostic is
 specifically tailored to be applied directly on baryon 
acoustic oscillation data, and depends on fewer cosmological observables than $Om$ in this case.       

Combining BAO information from the SDSS, WiggleZ \& 6dF surveys with SNIa 
information from
 the Union2.1 data
set, allows us to determine the three point diagnostic, $Om3$, at several independent redshift bins thereby
placing robust constraints on the nature of dark energy.

\section{The Om3 diagnostic reconstructed from SNIa and BAO data}
A key role in the determination of Baryon Acoustic Oscillations is played
by the `dilation-scale' distance \cite{eisenstein}
\beq
D_V(z) 
= \left [D(z)^2\frac{cz}{H(z)}\right ]^{1/3}~,
~~~{\rm where} ~~D(z) := \frac{D_L(z)}{1+z} = \int{\frac{c dz}{H(z)}}~,
\label{eq:DV}
\eeq
and we have used $D_L(z) = (1+z)^2D_A(z)$ to relate the luminosity distance $D_L$
to the angular size distance $D_A$ and also made the assumption that the universe is spatially
flat.

Other important parameters which can be extracted from BAO's include the acoustic parameter
\beq
A(z) = \frac{100 D_V(z)\sqrt{\Omega_{0m}h^2}}{cz}
\label{eq:A}
\eeq
and the ratio
\beq
d(z) = \frac{r_s(z_{\rm CMB})}{D_V(z)}
\eeq
where $r_s(z_{\rm CMB})$ is the sound horizon at the epoch when CMB photons decouple from baryons.

Equation (\ref{eq:DV}) allows us to relate the expansion history,
$H(z)$, to $D_V$ and $D$ as follows
\beq
H(z) 
= \frac{czD(z)^2}{D_V(z)^3}~.
\eeq
Independent measurements of $D_V$ and $D_L$ can therefore be used to
reconstruct $H(z)$, as discussed in \cite{shaf_clark}.

In this paper we shall demonstrate that the {\em ratio} of the Hubble
parameter at two redshifts can play a key role in
cosmological reconstruction.
To see this note that
\beq
H(z_i;z_j) := \frac{H(z_i)}{H(z_j)} =
\frac{z_i}{z_j}\left [ \frac{D(z_i)}{D(z_j)}\right ]^2\left [\frac{D_V(z_j)}{D_V(z_i)}\right ]^3~, 
\label{eq:fracH1}
\eeq
equivalently 
\beq
H(z_i;z_j) =
\left (\frac{z_j}{z_i}\right )^2\left [ \frac{D(z_i)}{D(z_j)}\right ]^2\left [\frac{A(z_j)}{A(z_i)}\right ]^3
= \frac{z_i}{z_j}\left [ \frac{D(z_i)}{D(z_j)}\right ]^2\left [\frac{d(z_i)}{d(z_j)}\right ]^3~,
\label{eq:fracH2}
\eeq
in other words, ratio's of BAO parameters
$D_V, A, d(z)$ are related to the ratio of the Hubble parameter.
From (\ref{eq:fracH2}) we find that $H(z_i;z_j)$ does
not depend either on $H_0, \sqrt{\Omega_{0m}h^2}$, or even the CMB parameter
$r_s(z_{\rm CMB})$.

We therefore arrive at the following important result: the value of
$D_L$ together with the value of {\em either of}
the
BAO paramaters 
$\lbrace D_V, A, d\rbrace$, determined at {\em three independent redshifts},
is sufficient to evaluate the $Om3$ diagnostic and define
a null test of the cosmological constant hypothesis ! 

This follows immediately
from (\ref{eq:om}) \& (\ref{eq:om3}), since, from the definition of $Om3$ in (\ref{eq:om3})
it follows that
\beq
Om3(z_1;z_2;z_3) = \frac{H(z_2;z_1)^2-1}{x_2^3-x_1^3}\bigg/
\frac{H(z_3;z_1)^2-1}{x_3^3-x_1^3}, ~~{\rm where} ~~x = 1+z, 
\label{eq:om3a}
\eeq
where $H(z_i;z_j)$ is determined using (\ref{eq:fracH1}) or (\ref{eq:fracH2}).
We should note that interchanging redshifts allows us to define different variants of $Om3$ ($z_1$, $z_2$ and $z_3$ should not be necessarily sorted in redshift). 
Large galaxy redshift survey's including SDSS, WiggleZ and 6dFGS
 have determined the BAO parameters
at several distinct redshift points \cite{percival,wigglez1,wigglez2,6dF}. 
In this paper we shall utilize these results 
to reconstruct $Om3$, thereby obtaining useful constraints
on the nature of dark energy. It should be stressed that
the model independent method which we advocate
does not rely on a knowledge of either the matter density $\Omega_{0m}$ or
the Hubble parameter $H_0$.
Its scope and prowess is therefore likely to dramatically 
improve as new data is added to
the burgeoning BAO inventory by dedicated upcoming surveys such as 
BOSS \cite{BOSS}, BigBOSS \cite{BigBOSS}, 
J-PAS \cite{JPAS} and Euclid \cite{Euclid}.

\begin{figure*}[!t]
\centering
\begin{center}
\vspace{0.7in}
\centerline{\mbox{\hspace{0.in} \hspace{2.1in}  \hspace{2.1in} }}
$\begin{array}{@{\hspace{-0.3in}}c@{\hspace{0.3in}}c@{\hspace{0.3in}}c}
\multicolumn{1}{l}{\mbox{}} &
\multicolumn{1}{l}{\mbox{}} \\ [-2.8cm]
\hspace{0.in}
\includegraphics[scale=0.45, angle=-90]{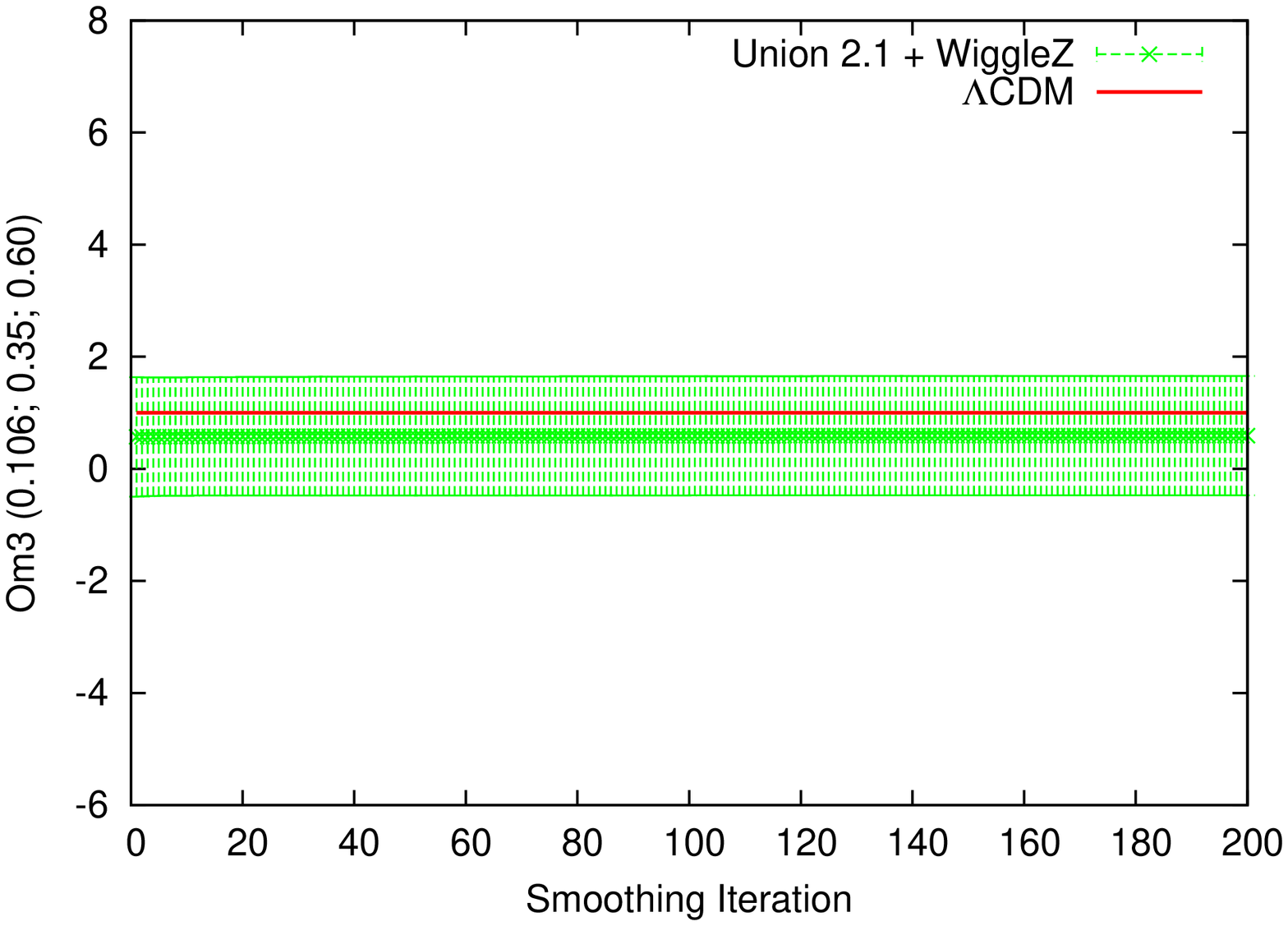}
\hspace{0.in}
\includegraphics[scale=0.45, angle=-90]{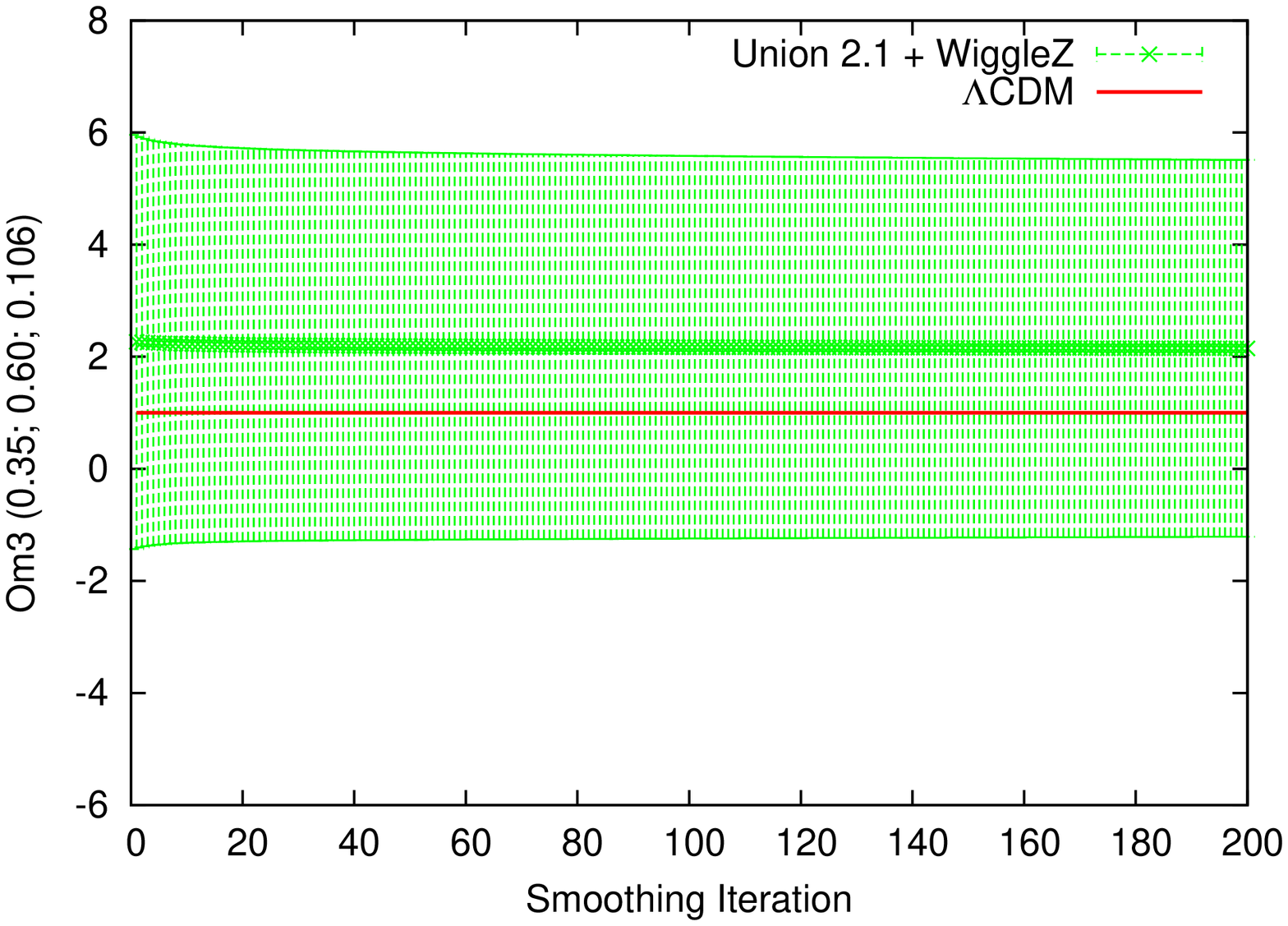}
\hspace{0.in}
\vspace{1.in}
\end{array}$
\vspace{-0.in}
$\begin{array}{@{\hspace{-0.3in}}c@{\hspace{0.3in}}c@{\hspace{0.3in}}c}
\multicolumn{1}{l}{\mbox{}} &
\multicolumn{1}{l}{\mbox{}} \\ [-2.8cm]
\hspace{0.in}
\includegraphics[scale=0.45, angle=-90]{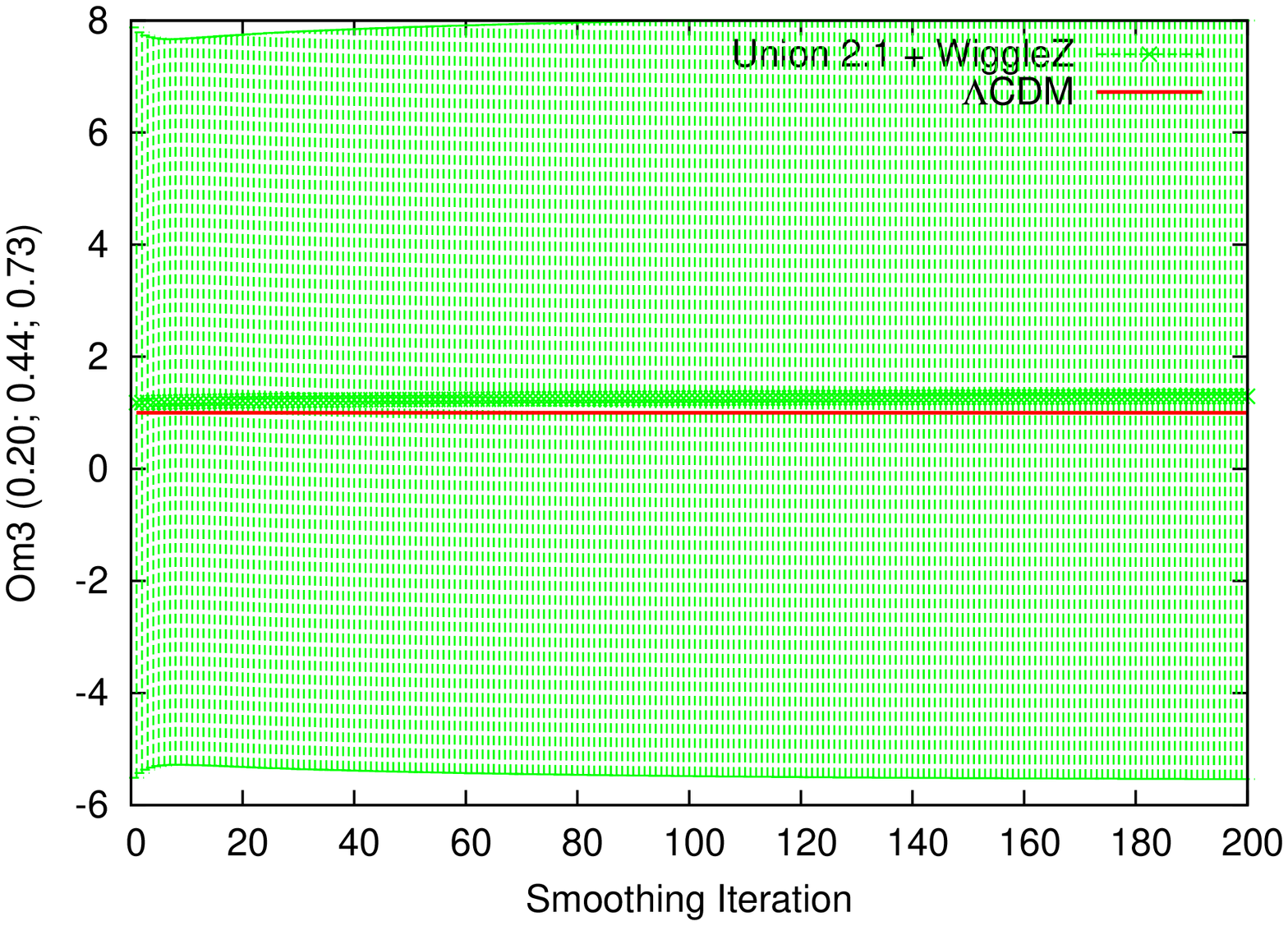}
\hspace{0.0in}
\includegraphics[scale=0.45, angle=-90]{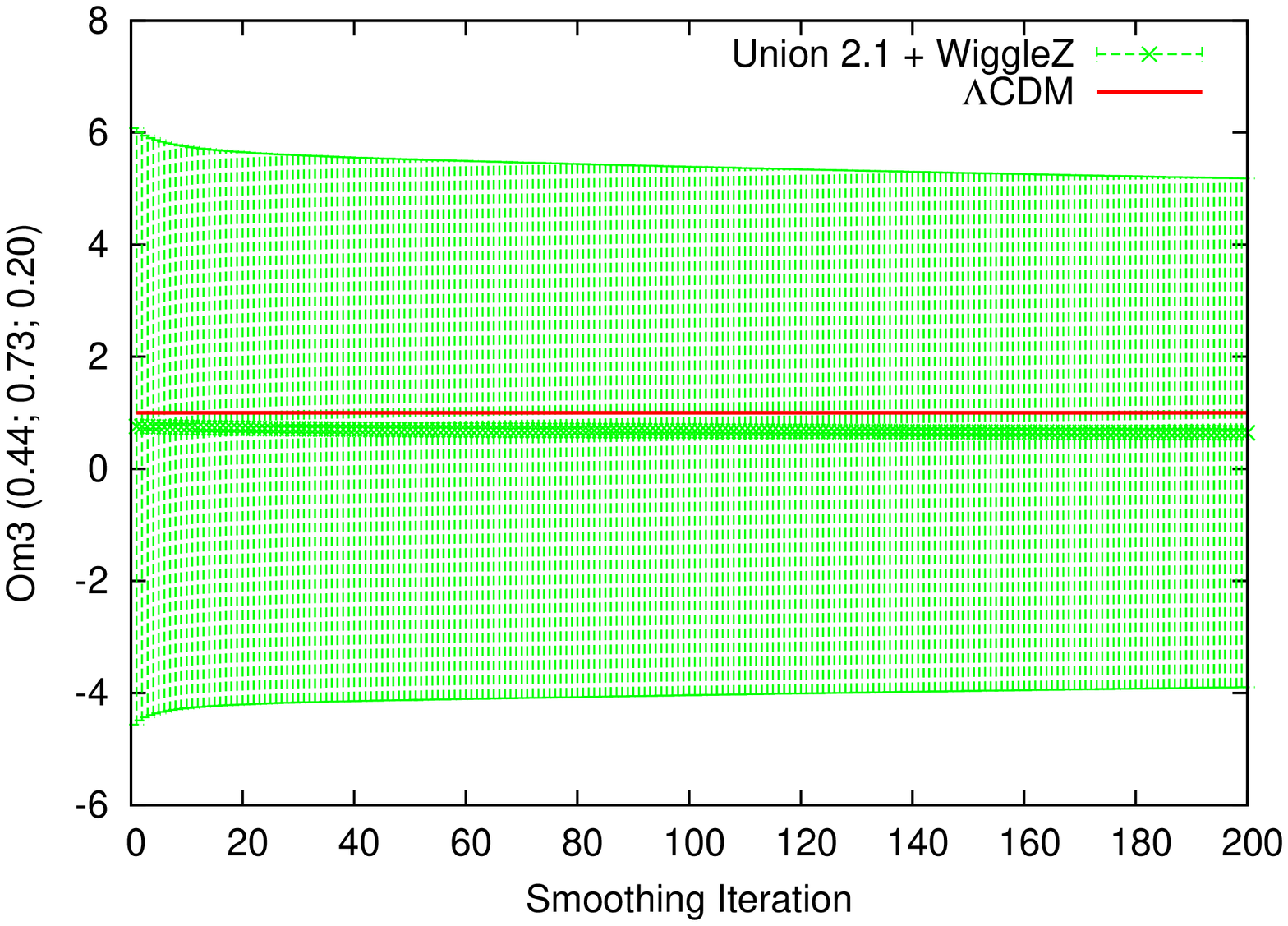}
\hspace{0.in}
\vspace{1.in}
\end{array}$
\vspace{-0.in}
$\begin{array}{@{\hspace{-0.3in}}c@{\hspace{0.3in}}c@{\hspace{0.3in}}c}
\multicolumn{1}{l}{\mbox{}} &
\multicolumn{1}{l}{\mbox{}} \\ [-2.8cm]
\hspace{0.in}
\includegraphics[scale=0.45, angle=-90]{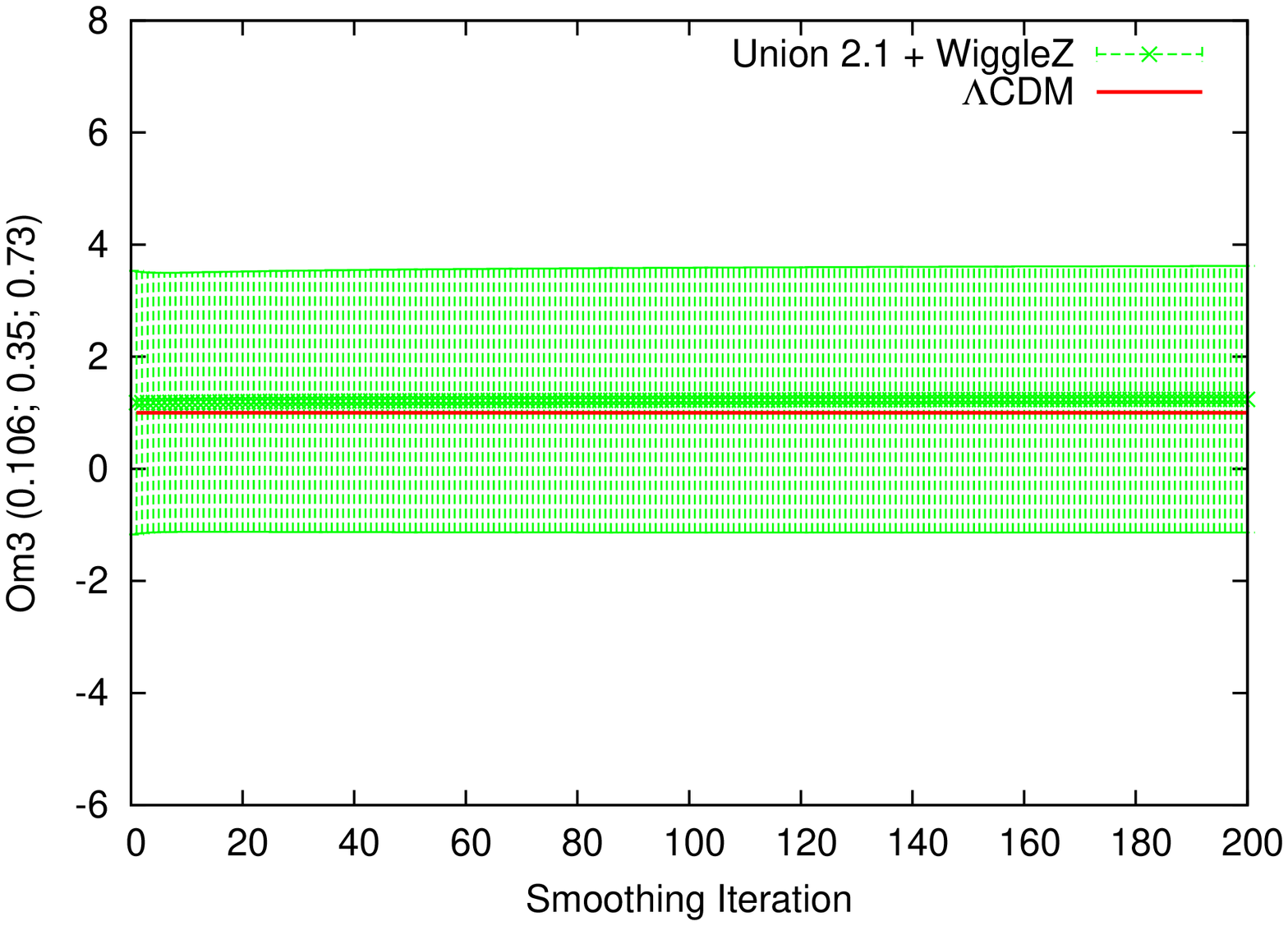}
\hspace{0.0in}
\includegraphics[scale=0.45, angle=-90]{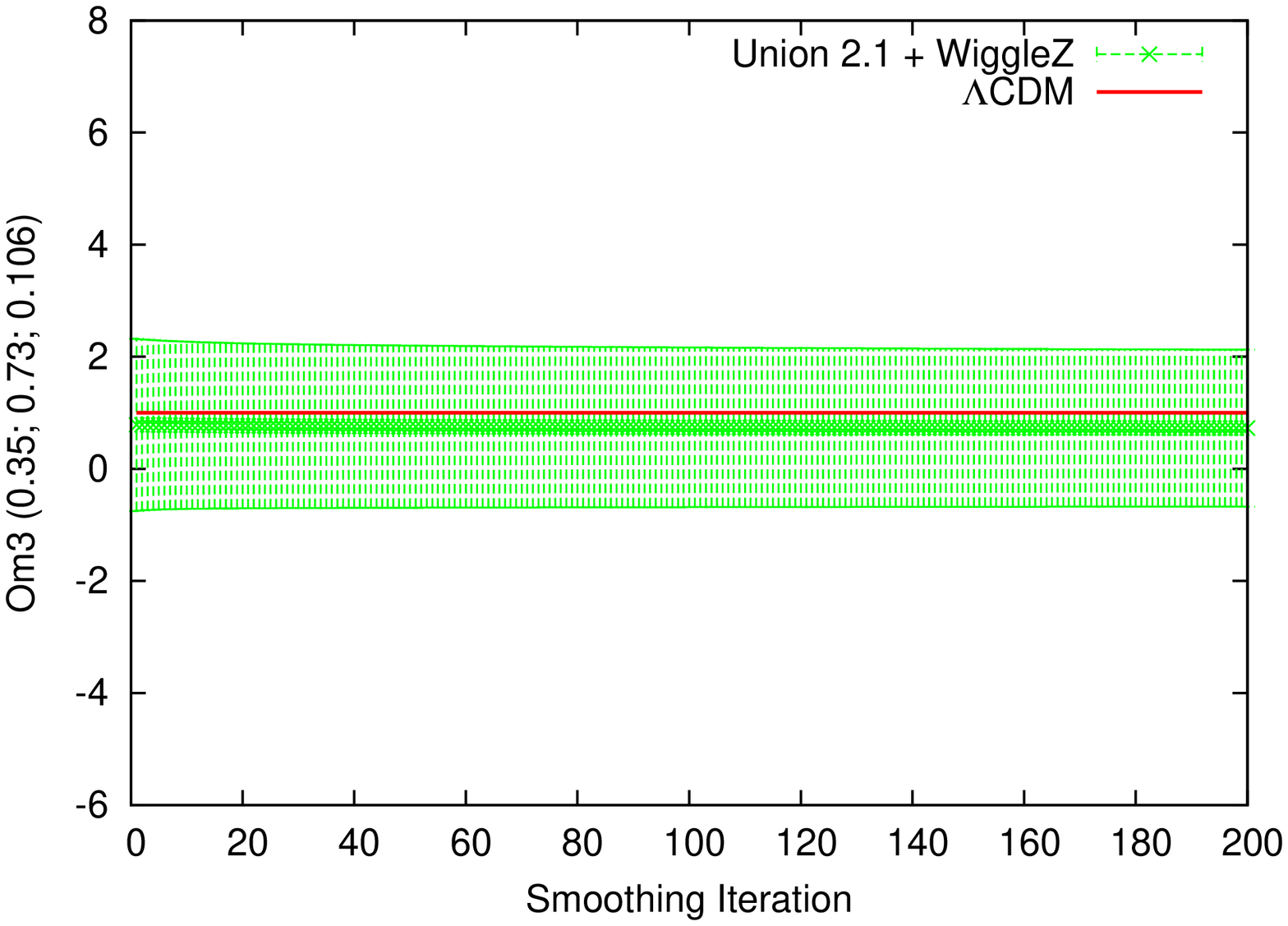}
\hspace{0.in}
\end{array}$
\vspace{-0.in}
\end{center}
\caption {\small Two variants of $Om3$ 
 determined using uncorrelated 
BAO measurements at three redshifts
in conjunction with the luminosity distance reconstructed from Union 2.1 
SNIa data. 
The redshift combinations $Om3(0.106,0.35,0.60)$ \& $Om3(0.35,0.73,0.106)$ shown in
the top left and bottom right panels have the smallest error bars and therefore
provide the tightest constraints on DE from current BAO+SNIa data. The large errorbars in the 
middle panels for $Om3$ are mainly due to the poor estimation of $d(z=0.44)$ (see Table 1). 
The spatially flat $\Lambda$CDM model is in good agreement with the data.
Note that these results do not rely on a knowledge of $\Omega_{0m}$, $H_0$,
or the distance to the last scattering surface.}  
\label{fig:WiggleZ}
\end{figure*}

\subsection{Cosmological Data Sets}

As we stressed above, in order to determine $Om3$
one only needs to know
the luminosity distance, $D_L(z_i)$, and {\em any one}
of the BAO parameters $D_V(z_i), A(z_i), d(z_i)$.
In this paper we shall use the Union 2.1 supernova data set
\cite{Union21} to reconstruct
$D_L$, and the SDSS DR7 \cite{percival}, WiggleZ \cite{wigglez1,wigglez2}
and 6dFGS \cite{6dF}
 determinations of the BAO parameters to determine $Om3$.
Below we briefly describe these data sets and our method of extracting $Om3$
from them.

\begin{figure*}[h]
\centering
\begin{center}
\vspace{0.7in}
\centerline{\mbox{\hspace{0.in} \hspace{2.1in}  \hspace{2.1in} }}
$\begin{array}{@{\hspace{-0.3in}}c@{\hspace{0.3in}}c@{\hspace{0.3in}}c}
\multicolumn{1}{l}{\mbox{}} &
\multicolumn{1}{l}{\mbox{}} \\ [-2.8cm]
\hspace{0.in}
\includegraphics[scale=0.45, angle=-90]{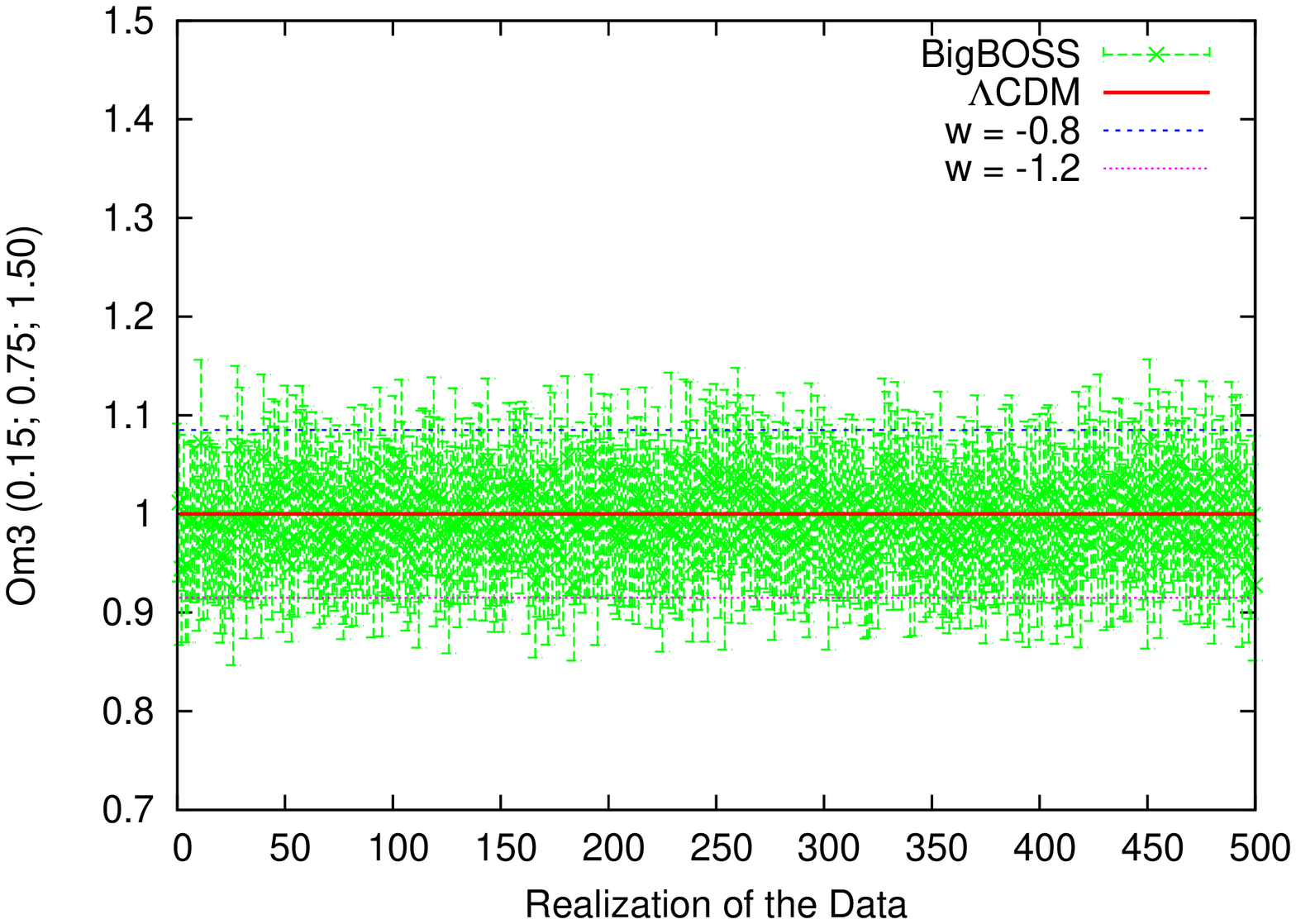}
\hspace{0.in}
\includegraphics[scale=0.45, angle=-90]{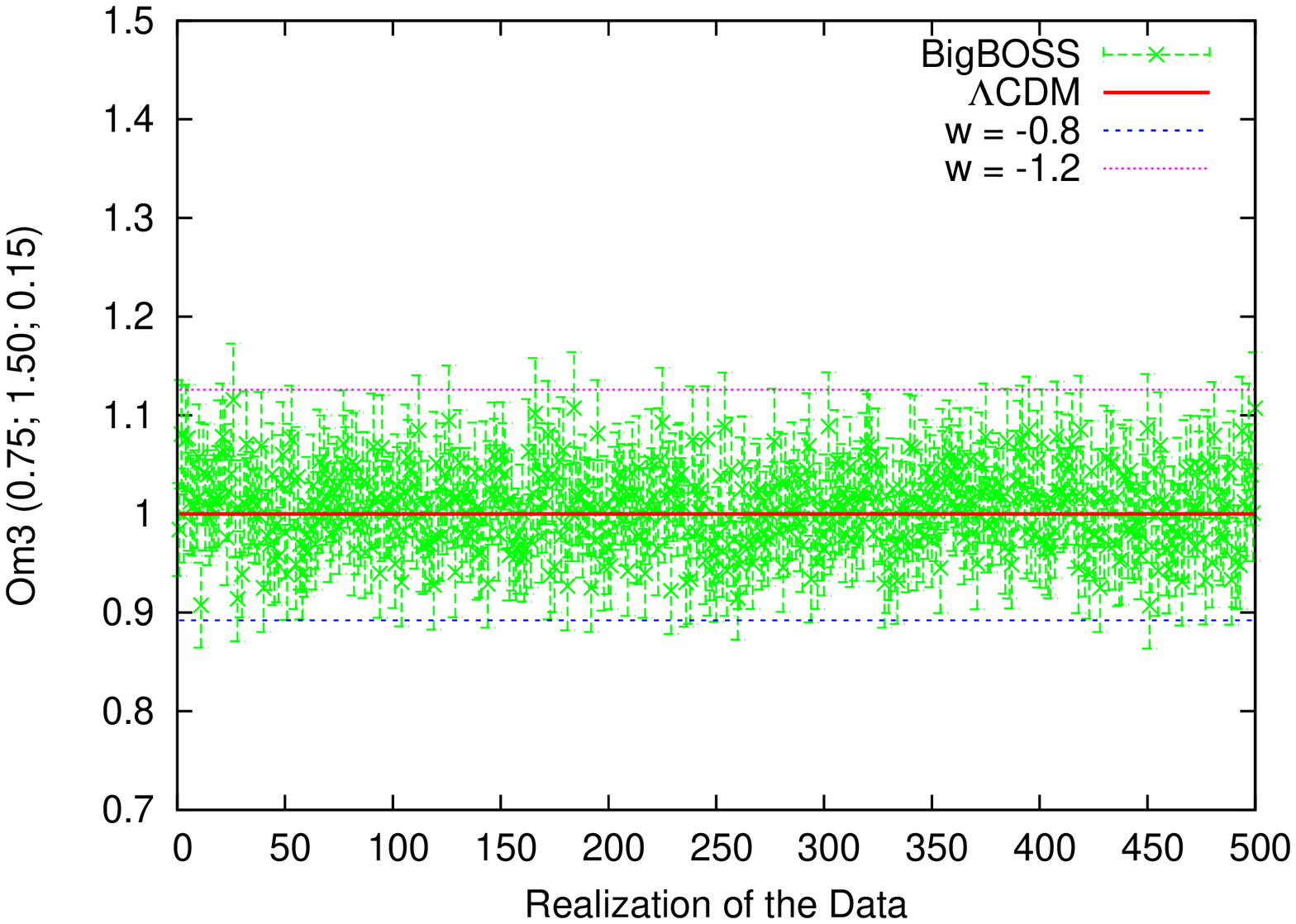}
\hspace{0.in}
\vspace{.in}
\end{array}$
\end{center}
\caption {\small Two variants of $Om3$ derived using simulated realizations of the 
BigBOSS experiment assuming a fiducial $\Lambda$CDM cosmology. Horizontal lines 
represent different dark energy models with (top-down) $w=-0.8$,
 $w=-1.0$, $w=-1.2$. $\Omega_{0m}=0.27$ is assumed for all models. 
Note that the determination of $Om3$ requires {\em minimal cosmological
assumptions} since one does not
require a background model to estimate $\Omega_{0m}$, $H_0$, or 
 the distance to the last scattering surface. 
Note also the difference in scale of the y-axis relative to figure \ref{fig:WiggleZ}.
}   
\label{fig:BigBOSS}
\end{figure*}

\begin{figure*}[h]
\centerline{\psfig{figure=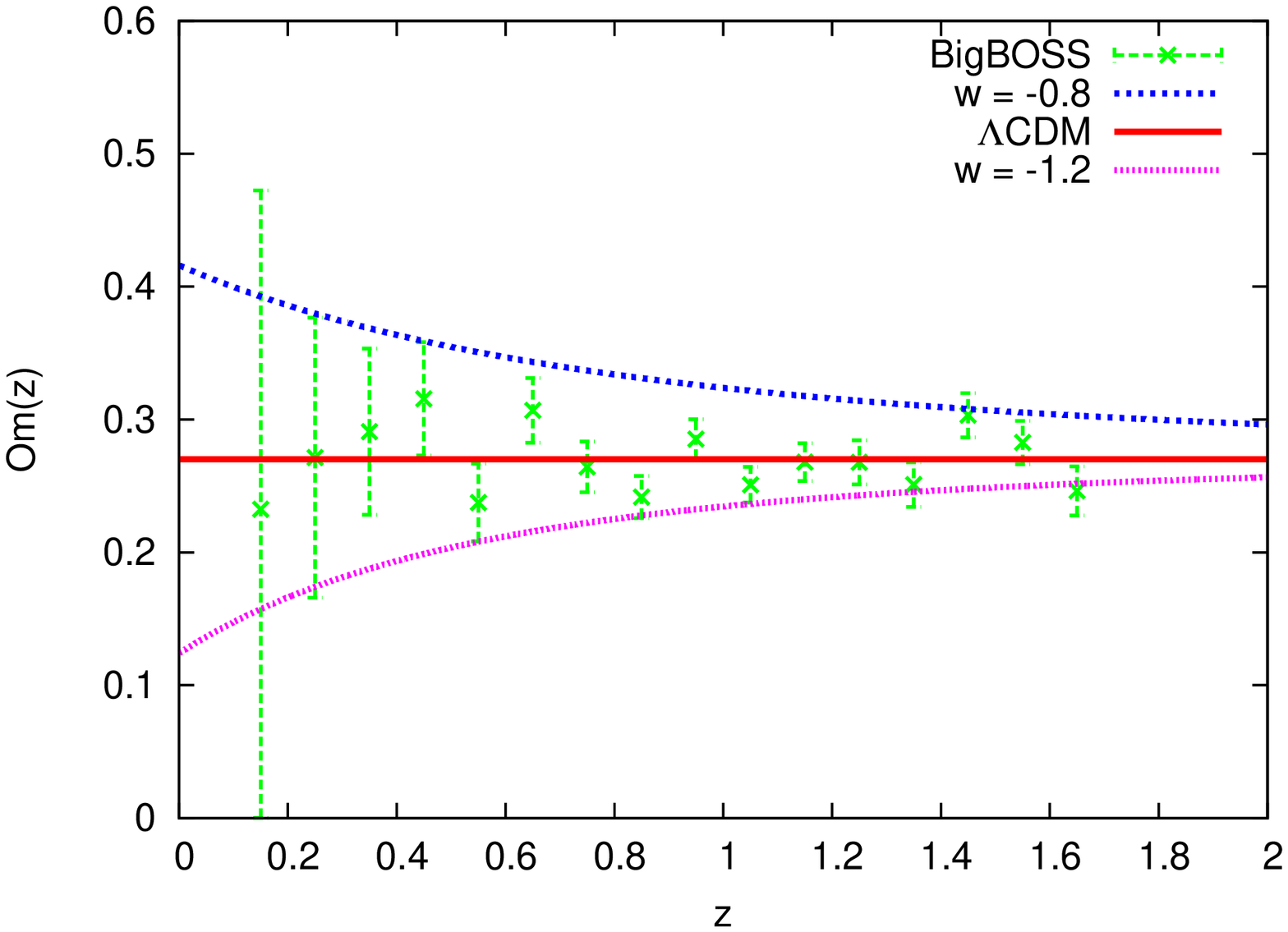,width=0.45\textwidth,angle=-90} }
\bigskip
\caption{\small The $Om$ diagnostic is reconstructed for a single simulated realization of the planned BigBOSS experiment assuming a fiducial $\Lambda$CDM cosmology.
A determination of $h(z)$, and hence
$Om(z)$, from future BAO experiments can clearly help
distinguish between rival 
models of dark energy. Note that this determination  is based on values of
$H_0$ with $2\%$ uncertainty expected by the time BigBOSS becomes operational.
}
\label{fig:Om_BigBOSS}
\end{figure*}


\begin{enumerate}

\item {\em Union 2.1 SNIa data.}
The Union 2.1 set \cite{Union21} 
consists of 580 type Ia supernovae sampling
the redshift range $0.015 \leq z \leq 1.414$. This data set includes 175 SNIa
at low redshifts $z < 0.1$.

Union 2.1 supernovae allow us to determine the luminosity distance $D_L(z)$ in a 
{\em model independent manner} following the efficient {\em smoothing ansatz}
 proposed in 
\cite{smooth}. Namely, a smoothed value for the luminosity distance, $D_L^S(z)$,
is constructed from the fluctuating `raw' value implied by data, $D_L(z)$, 
by smoothing
the latter using a low pass filter $F$ having a smoothing scale $\Delta$
\beq
D_L^S(z) = \int D_L(z')F(|z-z'|;\Delta) dz'~,
\eeq
where $F(|z-z'|;\Delta)$ is a variant of the Gaussian filter
$F_{\rm G} \propto \exp(-|{\bf x}-{\bf x'}|^2/2\Delta^2)$.
Specifically, one follows the iterative procedure \cite{smooth}
\ber
\label{eq:bg}
\ln D_L^S(z,\Delta)=\ln
\ D_L^g(z)+N(z) \sum_i \left [ \ln D_L(z_i)- \ln
\ D_L^g(z_i) \right] 
{\times} \ {\rm exp} \left [- \frac{\ln^2 \left
( \frac{1+z_i}{1+z} \right ) }{2 \Delta^2} \right ], &&\\
N(z)^{-1}=\sum_i {\rm exp} \left
[- \frac{\ln^2 \left ( \frac{1+z_i}{1+z} \right ) }{2 \Delta^2}
\right ]~, \hspace{2.8cm}&&\nonumber
\eer
where $D_L^g(z)$ is a `guessed' background model which is subtracted
from the data before smoothing, thereby ensuring that it is the noise
that is smoothed and not the luminosity distance !
The background model is upgraded at each iteration step via $D_L^g(z) \to D_L^S(z)$
and it is found that convergence is reached within a few steps, and that
cosmological reconstruction is quite insensitive to
the initial guess value
$D_L^g(z)$, which is reassuring. One might note that cosmological reconstruction
using the smoothing ansatz
 usually yields a better fit (improved $\chi^2$) when compared
with reconstruction
using parametric methods \cite{shaf_clark}.
The sensitivity of this method can be further enhanced by making the reconstruction
 error-sensitive
via the substitution $\left [ \ln D_L(z_i)- \ln
\ D_L^g(z_i) \right] \to \left [ \ln D_L(z_i)- \ln
\ D_L^g(z_i) \right]/\sigma^2_{D_L(z_i)}$ in (\ref{eq:bg});
see \cite{smooth,shaf_clark,arman} for more details,
and \cite{blake,shaf_clark,Crossing_rec} for other applications of the smoothing method. 

\item {\em Data from Baryon Acoustic Oscillations }

Baryon acoustic oscillations -- a relic of the
pre-recombination universe -- have been measured at 6 redshifts:
$z = 0.106,\, 0.2,\, 0.35,\, 0.44,\, 0.6$ and $0.73$. At these redshifts, the
`distilled BAO parameters' $A(z)$ and $d(z)$ have been determined to good accuracy,
with the BAO detection itself being at the level of $2-3\sigma$.
Table 1, which has been reproduced from \cite{wigglez2}, summarizes the BAO dataset. 

\begin{table}
\centering \caption{
BAO distances, from \cite{wigglez2}
}
\label{tab:margin} \footnotesize
\begin{center}
\begin{tabular}{c|c|c|c}
\hline
BAO sample&$z$&$d(z)$&$A(z)$\\
\hline
6dFGS&$0.106 $&$0.336\pm0.015$&$0.526 \pm 0.028$\\
&&&\\
SDSS&$0.2 $&$0.1905\pm0.0061$&$0.488 \pm 0.016$\\
&&&\\
SDSS&$0.35$&$0.1097\pm0.0036$&$0.484 \pm 0.016$\\
&&&\\
WiggleZ& 0.44 &$0.0916 \pm 0.0071$ &$0.474 \pm 0.034$\\
&&&\\
WiggleZ& 0.6 & $0.0726 \pm 0.0034$ & $0.442 \pm 0.020$\\
&&&\\
WiggleZ& 0.73 & $0.0592 \pm 0.0032$ & $0.424 \pm 0.021$\\
\hline
\end{tabular}
\end{center}
\end{table}

Not all of the BAO data in Table 1 is statistically independent. As pointed out in
\cite{wigglez2}, measurements belonging to the following redshift pairs
 are correlated:
$z = (0.2, 0.35), z = (0.44, 0.6), z = (0.6, 0.73)$,
the correlation coefficient being 0.337, 0.369 and 0.438, respectively.
By using a BAO parameter associated with one of the two redshifts 
belonging to a correlated pair, one has 3 relatively independent parameters in all, which is
precisely the correct number to determine $Om3$ !

In this paper we use $d(z)$ results from Table 1 associated with
 the following redshift triplets to determine $Om3$: $(0.106, 0.35, 0.6)$, $(0.2, 0.44, 0.73)$ and
$(0.106, 0.35, 0.73)$. 
Our main result, therefore, consists of finding
three independently obtained values for the 
$Om3$ diagnostic constructed from (\ref{eq:om3a})  \&  (\ref{eq:fracH2}) and using
 the BAO data in Table 1 jointly with Union 2.1 SNIa data.

\end{enumerate}

\subsection{Results}

In figure~\ref{fig:WiggleZ} we show results for two variants of $Om3$ using three uncorrelated measurements of the WiggleZ baryon acoustic oscillation survey. 
In the left column the results for the $(z_1;z_2;z_3)$ variant of $Om3$ are presented,
 while the right columns shows results for the $(z_2;z_3;z_1)$ variant. 
The spatially flat $\Lambda$CDM model ($Om3=1$ for all variants)
 is in good accord with the data. However, the quality of the data is not yet good enough 
to allow us to confront different cosmological models with data in a precise and,
 at the same time, model independent way. Propagation of errors results in
 large error bars on $Om3$ making
 it difficult to distinguish DE models from each other. 
It therefore appears that $Om3$ has the potential to be
 used as a model independent future probe of cosmological models when the quality of data 
improves significantly. In figure~\ref{fig:BigBOSS} we show
 the expected value of $Om3$ determined using
 500 simulated realizations of the BigBOSS experiment~\cite{BigBOSS}. 
One clearly sees that $Om3$ carries the potential to strongly discriminate between rival
 cosmological models with least a-priori assumptions being made about the early/late universe. 

While $Om3$ can be very useful for testing cosmological models with minimal a-priori assumptions, 
having reliable information from the early and late universe can help us to determine $h(z)$ and,
 knowing the latter, one can easily reconstruct the $Om$ diagnostic. In figure~\ref{fig:Om_BigBOSS} one 
realization of BigBOSS data together with expected future
measurements of $H_0$ with $2\%$ uncertainty is used to determine $Om(z)$. 
As we see from this figure, the $Om$ diagnostic provides
 a powerful means to discriminate between rival DE models.

\section{Conclusions}

In this paper we introduce a new null diagnostic customized for reconstructing the properties of 
dark energy from BAO data. $Om3$, as a 3 point diagnostic of dark energy,
 is closely related to the $Om$ diagnostic and follows the same general principles. 
$Om3$ is designed in a way that can be applied directly to BAO and supernovae data
in order to falsify concordance cosmology. This is done in a completely non-parametric way. 
The importance of $Om3$ lies in the fact that it does not rely on a knowledge of 
$\Omega_{0m}$, $H_0$, or the distance to the last scattering surface, which permits $Om3$ to falsify 
the $\Lambda$CDM model independently of these quantities. 
$Om3$ shares a common property with its cousin the $Om$ diagnostic, namely, the values of both 
$Om(z)$ and $Om3(z)$ stay pegged (at unity for $Om3(z)$ and $\Omega_{0m}$ for $Om(z)$) in an expanding concordance cosmology ($\Lambda$CDM).
This property of $Om3$ serves as a {\rm null test} of the cosmological constant hypothesis, since,
if observations do indicate that $Om3$ evolves with redshift, then this would imply
that $w \neq -1$ for the equation of state of dark energy.
We have shown that 
current BAO and supernovae data are in agreement with standard $\Lambda$CDM, however the uncertainties on 
$Om3$ are still pretty high. This is mainly due to the quality of available BAO data 
(current BAO data alone cannot put tight constraints on the cosmological quantities~\cite{Crossing_rec}) 
but future experiments, such as BigBOSS, can lead to much tighter determinations of $Om3$. 
Independence of $Om3$ from any a-priori 
assumptions of the early universe as well as values of $H_0$ and $\Omega_{0m}$ 
are important salient features of this null diagnostic of concordance cosmology ($\Lambda$CDM).

Finally we would like to highlight the main differences between the two null diagnostics of
$\Lambda$CDM -- $Om$ and $Om3$. 
 $Om$ is quite clearly a powerful null test of cosmological constant,
 and as it requires only a knowledge of $h(z)$,
 it seems straightforward to use $Om$ together with parametric or non-parameteric methods of reconstruction.
However deriving $h(z)$ directly from cosmological observables in a purely model independent and non-parameteric manner is not always an easy task. 
Though using supernovae data one can still derive $h(z)$ in a model 
independent manner \cite{smooth}, deriving $h(z)$ directly from large scale structure data seems to be 
more difficult~\cite{shaf_clark}. In this paper we have shown that $Om3$, in contrast to $Om$,
 is specifically tailored to be applied to baryon acoustic oscillation data (directly through BAO observables) 
and that, in this case, $Om3$  depends on a fewer number of cosmological observables than $Om$.

\section*{Acknowledgments}
A.S. thanks Chris Blake for useful discussions and for providing us with the early WiggleZ data. A.S. acknowledge the Max Planck Society (MPG), the Korea Ministry of Education, Science and Technology (MEST), Gyeongsangbuk-Do and Pohang City for the support of the Independent Junior Research Groups at the Asia Pacific Center for Theoretical Physics (APCTP). He also acknowledges the support of the World Class University grant R32-2009-000-10130-0 through the National Research Foundation, Ministry of Education, Science and Technology of Korea. A.A.S. acknowledges RESCEU hospitality as a visiting professor. He was also partially supported by the grant RFBR 11-02-00643 and by the Scientific Programme ``Astronomy'' of the Russian Academy of Sciences.

\end{document}